\documentclass[referee]{aa} 
\usepackage{graphicx}
\begin{document}
%
   \title{Bright metal-poor variables: why
``Anomalous'' Cepheids?}

   \author{F. Caputo\inst{1}, V. Castellani\inst{1,2},
S. Degl'Innocenti\inst{3,4},
   G. Fiorentino\inst{1,5}, M. Marconi\inst{6}}




   \offprints{}

   \institute{INAF - Osservatorio Astronomico di Roma, Via di Frascati 33,
I-00040 Monteporzio Catone, Italy\and INFN,
Sezione di Ferrara, via Paradiso 12, 44100\and Dipartimento di
Fisica, Universita' di Pisa , via Buonarroti 2, I-56127 Pisa,
Italy\and INFN, Sezione di Pisa, via Buonarroti 2, I-56127 Pisa,
Italy\and
Universita' ``Tor Vergata'', via della Ricerca Sci
entifica 1, 00133,  Roma
   \and
INAF Osservatorio Astronomico di Capodimonte,
Via Moiariello 16, I-80131 Napoli, Italy}

\date{Received .........; accepted ........}
\titlerunning{Why ``Anomalous'' Cepheids?}
\authorrunning{Caputo et al.}

\abstract{We investigate  the theoretical scenario concerning the
large sample of variables recently discovered in the dwarf,
metal-poor irregular galaxy Leo A, focusing the attention on the
"Anomalous" Cepheid phenomenon and its correlation with RR Lyrae
stars, Classical  and Population II Cepheids. To this purpose, we
make use of suitable  stellar and pulsation models to depict the
pulsational history of evolutionary structures  with metallicity
$Z$=0.0004, finding that He-burning pulsators are expected only
outside the mass interval $\sim$ 0.8-1.7$M_{\odot}$.  Stars
from $\sim$ 1.8 to 4$M_{\odot}$, a mass range including both {\it
Anomalous} and {\it Classical} Cepheids, populate with good
approximation a common $M_V$-log$P_F$ instability strip,
independently of the previous occurrence of a He flash  event,
with periods and luminosities increasing with the stellar mass and
with a lower luminosity level $M_{V,LE}\sim -$0.5 mag, as observed
in Leo A. The  class of less massive pulsators ($M<$
0.8$M_{\odot}$, namely {\it RR Lyrae} stars and {\it Population II
Cepheids}) populate a distinct instability strip, where the
magnitudes become brighter and the periods longer when decreasing
the pulsator mass. The dependence on metal content of this
scenario has been investigated over the range $Z$=0.0002 to 0.008.
One finds that the edges of the pulsational strip for the more
massive class of pulsators appear independent of metallicity, but
with the minimum mass of these bright pulsators which decreases
when decreasing the metallicity, thus decreasing the predicted
minimum  luminosity and period. Comparison with data for Cepheids
in Leo A and  in the moderately metal rich extragalactic stellar
system Sextans A discloses an encouraging agreement with the
predicted pulsational scenario. On this basis, we predict that in
a stellar system where both RR Lyrae stars and Cepheids are
observed, their magnitude difference may help in constraining both
the metal content and the distance. The current classification of
metal-poor Cepheids is shortly discussed, advancing  the
suggestion for an updated terminology abreast of the current
knowledge of stellar evolution.

 }
  \maketitle

\section{Introduction}
Even though the majority of metal-poor radial pulsators belongs to
the RR Lyrae  class, other kinds of variables are observed both in
Galactic globular clusters and  in extragalactic metal-poor
stellar systems. According to the current nomenclature, among the
objects brighter than RR Lyrae stars one finds Population II
Cepheids (P2C) and Anomalous Cepheids (AC), the former with
periods from $\sim$ 1 to $\sim$ 25 days, the latter from $\sim
0.3$ to $\sim$ 2 days. Both these variables are interpreted in
terms of central He-burning structures, as RR Lyrae stars are, but
being either less massive (P2Cs) or more massive (ACs ) than RR
Lyrae stars with similar metal content.

Investigations based on the predictions of stellar evolution and
radial pulsation models have already shown that P2Cs originate
from hot, low-mass Zero Age Horizontal Branch (ZAHB) stars
evolving from large effective temperatures towards the Asymptotic
Giant Branch (AGB), crossing the instability strip at a luminosity
larger than the RR Lyrae level (see, e.g., Bono et al 1997a and
references therein, as well as the recent review by Wallerstein
2002).  Since the pulsation period $P$ increases with increasing
luminosity and/or with decreasing mass, one predicts that these
bright low mass pulsators should have longer periods than RR Lyrae
stars, as observed.

Concerning ACs, it has been shown  that at metal contents $Z\le$
0.0004 and for not-too-old ages ($\le$ 2 Gyr) the effective
temperature of  ZAHB models, which normally decreases with
increasing the mass, reaches a minimum at log$T_e\sim$ 3.74
($Z$=0.0001) or $\sim$ 3.72 ($Z$=0.0004) around
1-1.2$M_{\odot}$(see Castellani \& Degl'Innocenti 1995, Caputo \&
Degl'Innocenti 1995 and references therein). By further increasing
the mass over this value, both the luminosity and effective
temperature of the ZAHB structure increase, producing a "ZAHB
turn-over" and an "upper horizontal branch" which intersects the
instability strip again at a luminosity larger than the RR Lyrae
level (see Bono et al. 1997b). In this case the effect of the
higher luminosities is somehow "balanced" by the larger masses,
and consequently these bright massive pulsators will show periods
that are not significantly longer than those typical of RR Lyrae
stars, in agreement with the observed  behavior for ACs. On this simple basis,
one can understand why ACs and P2Cs appear to obey distinctive
Period-Luminosity relations and why, at a given luminosity, ACs
have shorter periods than P2Cs. It turns out that, at a given
period, observed ACs appear more luminous than P2Cs, a feature
which is at the origin of their supposed "anomaly".

ACs are very rare in globular clusters, whereas they have been
found in  all dwarf spheroidal galaxies that have been surveyed for
variable stars (see, e.g., Mateo 1998, Siegel \& Majewski 2000,
Pritzl et al. 2002, Da Costa, Armandroff \& Caldwell 2002,
Dall'Ora et al. 2003, Baldacci et al. 2004). Recently, Dolphin et
al. (2002, hereafter D02) reported the results of a search for
short-period ($P \le$ 2 days) variables in Leo A, a Local Group
dwarf irregular galaxy characterized by a very low metal abundance
($Z\sim$ 0.0004). According to D02, eight candidate RR Lyrae stars
have been found, with a mean magnitude of $<V>$=25.10$\pm$0.09
mag, suggesting the presence in this galaxy of an old ($\sim$
10-11 Gyr) stellar population. The same study reports the
discovery of several (84) variables brighter than $V$=24.5 mag and
with periods between $\sim$ 0.4 and 2 days. Although these
properties correspond to the well-recognized behavior of  ACs, the
authors suggest that, rather than being indicative of this class
of variables, they appear as a natural extension to low
metallicity of "classical" (i.e. metal-intermediate) short-period
Cepheids, such those observed in the Large and Small Magellanic
Clouds ($Z\sim$ 0.008 and $\sim$ 0.004, respectively).

In this paper we will show that ACs do represent indeed the
natural extension of  classical Cepheids to lower metallicities
and masses. To this end, in the following we will rely on suitable
sets of stellar evolution and pulsation models to discuss the
"Anomalous" Cepheid phenomenon and its correlation with RR Lyrae
stars, Population II and Classical Cepheids. The theoretical
scenario is presented in Section 2, while Section 3 discusses  the
predicted Period-Luminosity distribution when $Z$=0.0004. Section
4 deals with the dependence on the assumed metal content and the
comparison with observed data is given in Section 5. As a
concluding remark, in the last section we direct the reader's
attention to the evidence that the current classification might be
misleading in some respects, perhaps requiring an updated
terminology based on the current improved knowledge of stellar
evolution.

\section{The theoretical scenario}

Radial pulsation is a phenomenon appearing only in selected
evolutionary phases. Given a metal abundance, for each given
stellar mass and luminosity there is a maximum (blue edge) and a
minimum effective temperature (red edge) for the onset of the
pulsation instability. Varying the luminosity, for each given mass
one has the  so called "instability strip" which crosses the HR
diagram from the higher temperatures and lower luminosities
towards lower temperatures and higher luminosities. If and when
the evolving star intersects "its" mass-dependent instability
strip, it goes pulsating with a period strictly constrained by its
mass, luminosity, and effective temperature.

With respect to static stars, radial pulsating structures present
the undeniable advantage of at least an additional observable, as
given by the pulsation period, which, besides being unaffected by
distance and reddening, yields relevant constraints on the stellar
structural parameters. Combining evolution and pulsation models
one can predict the behavior of radial pulsators occurring in the
different evolutionary phases, providing a theoretical frame for a
sound interpretation of the different classes of observed
variables. In this way, we have recently presented a detailed
analysis of RR Lyrae stars in the Galactic globular cluster M3
(see Marconi et al. 2003).

 To discuss anomalous and classical Cepheids in Leo A, we will make
use of the pulsation models with $Z$=0.0001 and 0.0004 recently
computed by Marconi, Fiorentino \& Caputo (2004) for mass in the
range 1.3-2.2$M_{\odot}$ and luminosity log$L/L_{\odot}$=1.82 to
2.28, implemented with a set of 4$M_{\odot}$ models at $Z$=0.0004
and log$L/L_{\odot}$=3.5. By comparison with similar models
presented by Bono et al. (2002) for $Z$=0.008 over the mass range
3-5$M_{\odot}$, one finds that all over the range $Z$=0.0001 to
0.008 the blue (FOBE) and red (FRE) edges of the instability strip
follow the two relations

$$\log T_e(FOBE)=3.925(\pm 0.008)-0.052\log L+0.042\log M-0.006\log Z\eqno(1a)$$
$$\log T_e(FRE)=3.876(\pm 0.008)-0.065\log L+0.058\log M-0.006\log Z,\eqno(2a)$$
\noindent where $M$ and $L$ are in solar units, whereas the period
of fundamental (F) pulsators is given by

$$\log P_F=10.925(\pm 0.005)+0.818\log L-0.616\log M-3.309\log T_e+0.012\ log Z\eqno(3a)$$

 Moreover, in order to compare the predicted behavior of these
relatively massive pulsators  with that of the less massive RR
Lyrae and P2C variables, pulsation models with metal content
$Z$=0.0001-0.006, $M$=0.58-0.80$M_{\odot}$ and
log$L/L_{\odot}$=1.51 to 1.91  have been  considered, for which it
has already   been found (Marconi et al 2003, Di Criscienzo, Marconi \&
Caputo 2004)

$$\log T_e(FOBE)=3.970(\pm 0.005)-0.057\log L+0.094\log M\eqno(1b)$$
$$\log T_e(FRE)=3.957(\pm 0.010)-0.102\log L+0.073\log M\eqno(2b)$$
$$\log P_F=11.039(\pm 0.005)+0.833\log L-0.651\log M-3.350\log T_e+0.008\ log Z\eqno(3b)$$
\noindent
 In all cases one finds that first-overtone (FO)
pulsators can be "fundamentalised" by adopting
log$P_F$=log$P_{FO}$+0.13.

 All the pulsation models we are referring to  rely on the use of a
(same) nonlinear convective hydrodynamical code, an approach which
is of central importance to gain reliable information on the
boundaries for  fundamental or first overtone pulsation. The
inadequacy of linear computations in this respect has been indeed
already discussed by Bono et al. (1999). However, here we notice
that even in the linear approach  (see, e.g., Chiosi, Wood \&
Capitanio 1993), one predicts a significant dependence on the
pulsator structural parameters (mass and luminosity), at variance
with the hypothesis of a constant value (log$T_e$(FRE)=3.73) as
recently adopted by Cordier, Goupil \& Lebreton (2003).

 Such a pulsational scenario has been eventually combined with
canonical (i.e., no mass-loss, no overshooting) evolutionary
models with $Z$=0.0004 from the "Pisa Evolutionary Library"
(http://gipsy.cjb.net). All models cover the major phases of both
H and He burnings (see Cariulo, Degl'Innocenti \& Castellani 2003
for details). Following the procedure presented by Bono et al.
(1997 a,b) and Marconi et al. (2003), one can depict the
pulsational history of each evolutionary model (i.e., for each
given mass and original composition) by investigating the
evolutionary paths in terms of the difference between the actual
effective temperature and the predicted values at FRE and FOBE at
the various luminosities.

The straightforward consequence of this procedure is shown in Fig.
1 and Fig. 2, where evolutionary sequences with mass 0.8 to
4.0$M_{\odot}$ are plotted in a  "pulsational" HR diagram, where
effective temperatures are scaled, for each mass and luminosity,
to the appropriate FRE and FOBE, respectively.  For each given mass
the predicted pulsators are given by models with effective
temperature falling between the blue and red edges of the
pulsation region (FOBE $\ge$ log $T_e \ge$ FRE).

\begin{figure}
\includegraphics[width=10cm]{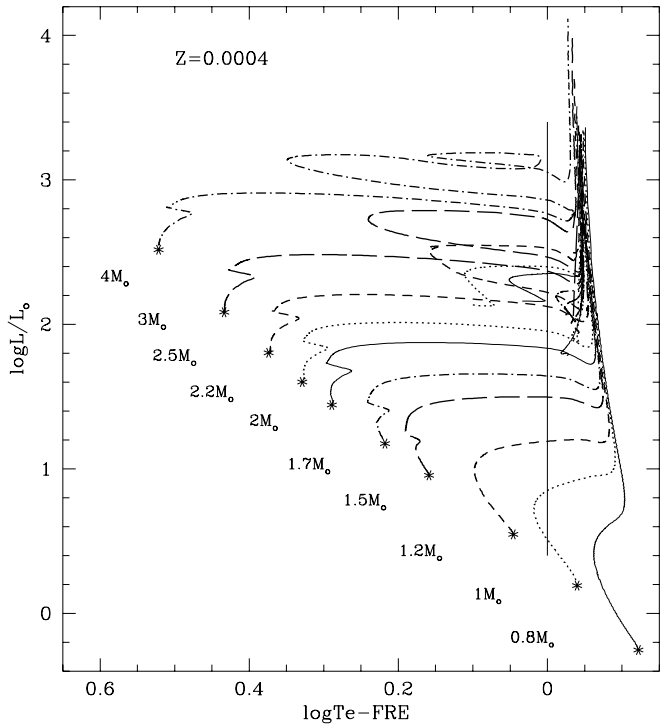}
\caption{Evolutionary tracks from Main Sequence to Asymptotic
Giant Branch at $Z$=0.0004 and for selected stellar masses. The
effective temperature is scaled to the predicted red edge (FRE) of
the pulsation region. Asterisks mark the MS models for the various
labelled masses.}
\end{figure}

\begin{figure}
\includegraphics[width=10cm]{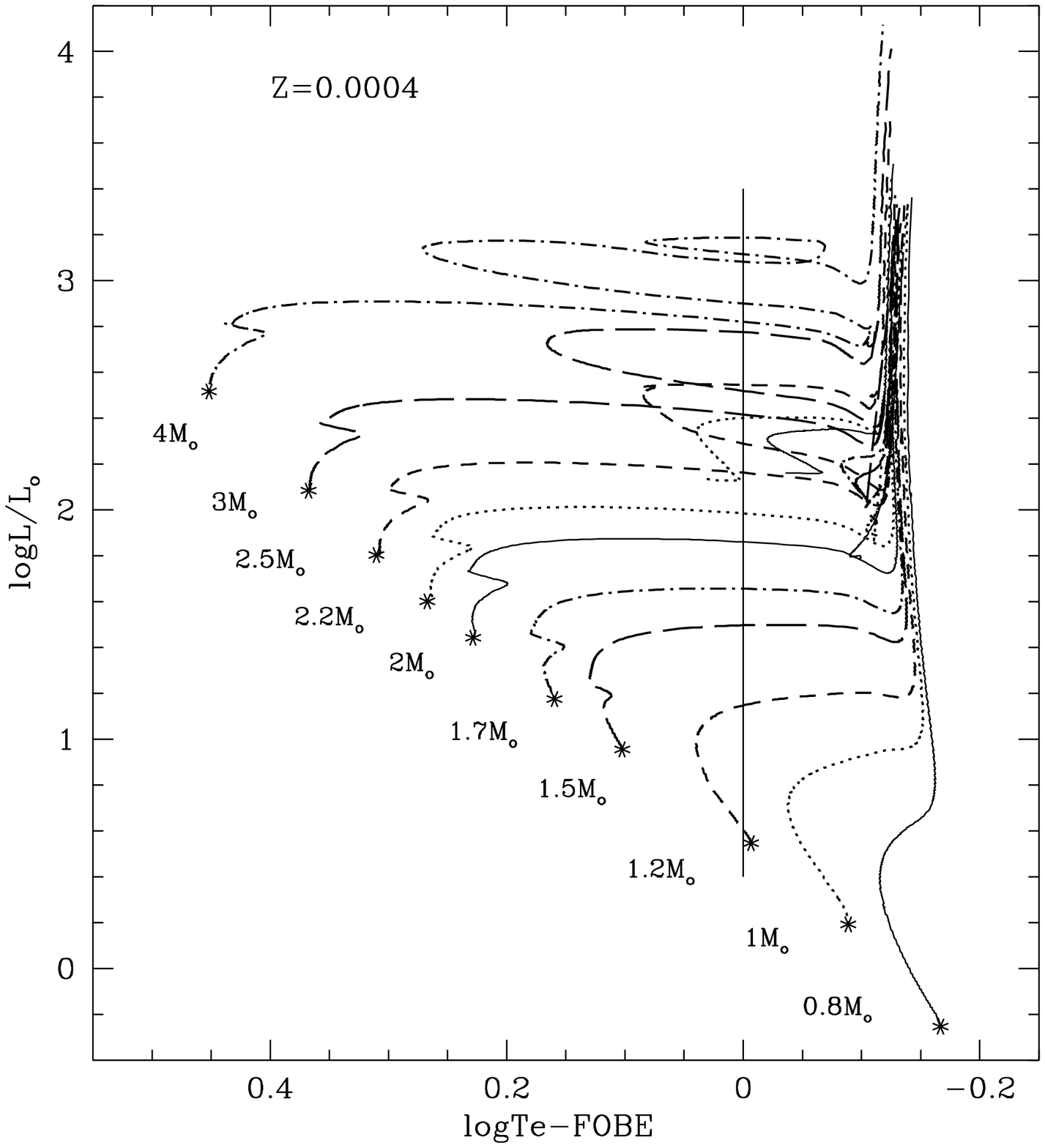}
\caption{As in Fig. 1, but with the effective temperature scaled
to the predicted blue edge (FOBE) of the pulsation region.}
\end{figure}

Inspection of both Fig. 1 and Fig. 2 provides a general overview
on the connection between pulsation and evolution. In particular,
varying the mass, one derives that:
\begin{enumerate}

\item   for masses smaller than $\sim$ 1.0$M_{\odot}$
  the whole evolution  proceeds at  effective temperature lower than FRE,
    and no variables are expected, unless mass-loss is driving
    the structures into the RR Lyrae instability region (see later);

\item   for masses from $\sim$ 1.0$M_{\odot}$ to $\sim$ 1.7$M_{\odot}$ the
        effective temperature of  He-burning models ("blue loops") is cooler
        than FRE, so that one would expect variables only during the
    short-lived H-burning phase when the stars evolve redward after
    central hydrogen exhaustion;

\item   for evolving stars with mass
        larger than 1.7$M_{\odot}$ crossing
        of the instability
        strip occur both in the phase following the
    exhaustion of central hydrogen and
        during central He-burning.
\end{enumerate}

\noindent  Thus, if mass-loss is neglected,  central He-burning
pulsators with $Z$=0.0004 are expected only from stars more
massive than 1.7$M_{\odot}$ and brighter than log$L/L_{\odot}\sim$
2.15.

 The effect of mass-loss and the consequent appearance of low-mass
central He-burning variables (i.e. RR Lyrae stars and P2Cs)
deserve further comments. For masses $M>$ 1.0$M_{\odot}$  the
effects of mass loss are substantially irrelevant : even losing
0.3$M_{\odot}$ by a 1.7$M_{\odot}$ progenitor
($M_c$=0.467$M_{\odot}$) will yield a 1.4$M_{\odot}$ He-burning
structure which is only slightly fainter than the 1.4$M_{\odot}$
model without mass-loss ($M_c$=0.477$M_{\odot}$, see Castellani \&
Degl'Innocenti (1995) for details) In contrast, a mass loss of
$\sim$ 0.1-0.2$M_{\odot}$ in less massive progenitors has dramatic
consequences, moving  the newly born He-burning structures to
large effective temperatures, along the ZAHB locus, so that they
will successively cross the instability strip.

All this is summarized in Fig.3, where we report  an expanded
portion of Fig. 2  showing the post He-flash, central He-burning
paths of structures with mass from 2.0 to 0.8$M_{\odot}$ in the
absence of mass loss (solid lines), compared with the predicted
boundaries of the instability region (vertical lines).The same
figure shows also the evolution from the Zero Age Horizontal
Branch (ZAHB: dashed line) of structures with mass
0.6-0.75$M_{\odot}$ (dotted line) originated from a
0.8$M_{\odot}$ progenitor.

\begin{figure}
\includegraphics[width=10cm]{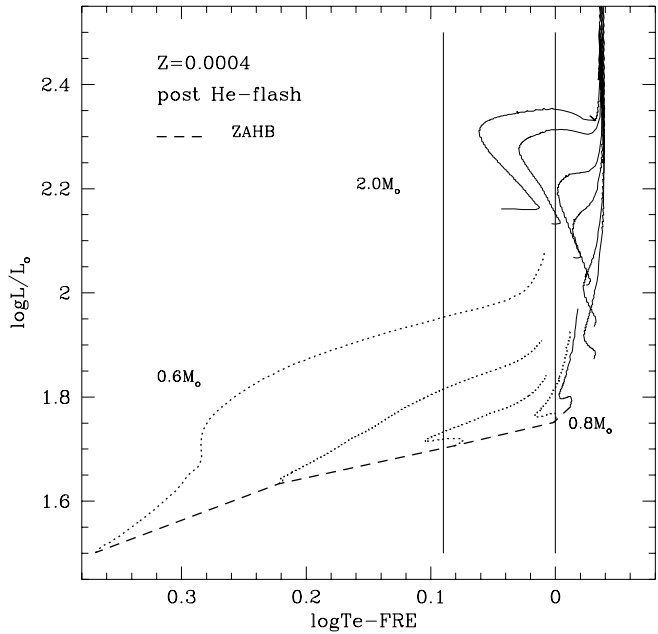}
\caption{Enlarged portion of Fig. 3 showing
the post He-flash,  central He-burning evolutionary tracks of
stellar structures with mass 0.8, 1.0, 1.2, 1.5, 1.7, 1.9 and
2.0 $M_{\odot}$ in absence of mass-loss during the RGB phase
(solid lines). The dotted lines depict the evolution from the ZAHB
(dashed line) of models with mass 0.60, 0.65, 0.70 and
0.75$M_{\odot}$ originating from  a 0.8$M_{\odot}$ progenitor.}
\end{figure}

 As expected, inspection of the figure shows that the predicted
luminosity of low mass ($<0.8M_{\odot}$) pulsators increases with
decreasing the mass, at variance with the mass-luminosity relation
followed by the more massive ($>1.7M_{\odot}$) structures. As a
conclusion, we can expect that

 a) with $Z$=0.0004 no central He-burning pulsators are expected
in the mass range $\sim$ 0.8-1.7$M_{\odot}$;

b) outside of this forbidden range,  more massive or less massive
pulsators should follow distinctive period-luminosity relations as
a consequence of their opposite mass-luminosity relations.

\section{The Period-Luminosity distribution}

By selecting the models falling within their instability strip,
one can now derive the predicted period-magnitude diagram of
pulsators with $Z$=0.0004 and various masses. To this purpose, we
used bolometric corrections by Castelli, Gratton \& Kurucz (1997)
to get absolute visual magnitudes, while fundamental periods $P_F$
are estimated for each given mass, luminosity and effective
temperature by means of the  relations given in the previous
section.

Figure 4 shows the predicted $M_V$-log$P_F$ diagram of central
He-burning pulsators at the labelled masses. One finds that stars from 1.9 to
4$M_{\odot}$ undergoing their radial pulsation phase define a
nearly unique instability strip,  independently of the occurrence
of a quiet  ($\le$ 2.1$M_{\odot}$) or a flashing ($>$
2.1$M_{\odot}$) He ignition. When increasing the mass, the
pulsators become generally brighter and have longer periods. As
shown in the figure, for these massive central He-burning
pulsators with $Z$=0.0004 one can predict a lower luminosity limit
of $M_{V,LE}\sim -$0.5 mag, as well as fundamental periods as
short as log$P_F\sim -$0.3. Concerning less massive pulsators
evolving off their ZAHB ($M<$ 0.8$M_{\odot}$, see Fig. 3), Fig. 4
shows that they populate a distinct instability strip, where the
magnitude becomes brighter and the period longer with decreasing
the pulsator mass.

\begin{figure}
\includegraphics[width=10cm]{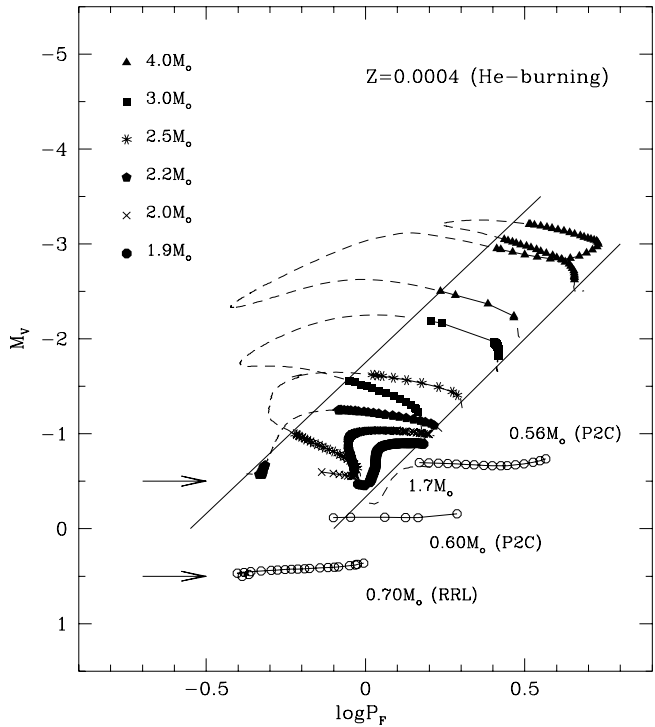}
\caption{Predicted absolute visual magnitude $M_V$ versus period
$P_F$ for He-burning fundamental pulsators with $Z$=0.0004 and
selected masses, in absence of mass-loss (solid symbols). The
portion of the tracks out of the pulsation region is drawn with a
dashed line. The solid lines show the edges of the predicted
$M_V$-log$P_F$ distribution, while the upper arrow marks the
predicted absolute magnitude $M_{V,LE}\sim -$0.5 mag of the lower
envelope of the pulsator distribution.  Open symbols illustrate
the behavior of the predicted RR Lyrae (RRL) and Population II
Cepheids (P2C) as caused by an efficient mass-loss. The lower
arrow shows the predicted absolute magnitude of RR Lyrae stars.}
\end{figure}

By inspection of the range of predicted (fundamental) periods  and
magnitudes,  in the pulsators with mass $\sim$ 1.9-2.2$M_{\odot}$
one can easily recognize typical features of observed ACs  in
dwarf spheroidal galaxies (see Marconi et al.  2004), while the
0.7$M_{\odot}$ and less massive pulsators should represent RR
Lyrae stars and Population II Cepheids, respectively. Note that
the predicted luminosity of RR Lyrae stars ($M_V$(RRL)$\sim$ 0.5
mag) is only $\sim$ 1 mag fainter  than the lower envelope of the
massive pulsator distribution ($M_{V,LE}\sim -$0.5 mag), whereas
the visual magnitude of the predicted P2C pulsators  can be even
brighter than $M_{V,LE}$. However, at a given period one finds ACs
brighter than P2C stars, as observed. As a whole, the results
plotted in Fig. 4 disclose that post-He flash structures with mass
$\sim$ 1.9-2.2$M_{\odot}$ (i.e, AC candidates) are the natural
extension to fainter magnitudes of the more massive central
He-burning pulsators of similar metal content.

\section{The effect of metallicity}

 The effect of metallicity on the  pulsational scenario has been
investigated  by repeating the above procedure but for the
selected metallicity values $Z$=0.0002, 0.0006, 0.001, 0.004 and
0.008, using again canonical evolutionary tracks from the Pisa
Evolutionary Library. As an example, the left panels in Fig. 5
show that the minimum mass for the occurrence of massive central
He-burning pulsators is passing from 1.9$M_{\odot}$ ($Z$=0.0004)
to $\sim$ 3.0$M_{\odot}$ if $Z$=0.004 and to 3.6$M_{\odot}$ when
$Z$=0.008. Correspondingly the lower luminosity level is
increasing from $M_V\sim -$0.5 mag to $M_V\sim -$1.3 mag
($Z$=0.004) and  $M_V\sim -$1.5 ($Z$=0.008).

\begin{figure}
\includegraphics[width=10cm]{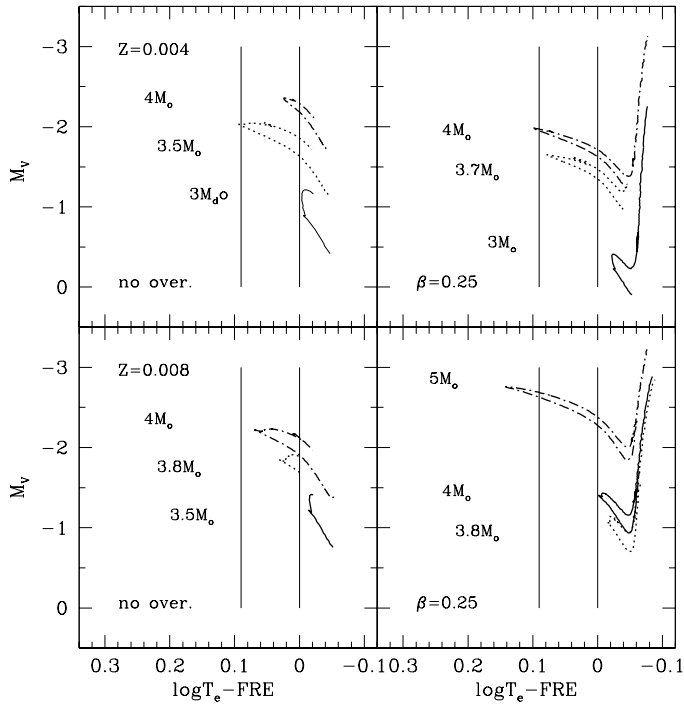}
\caption{Central He-burning tracks without (left panels) and with
overshooting (right panels) at the labelled metal content. The
effective temperatures are scaled to the predicted FRE. The solid
lines depict the pulsation region.}
\end{figure}

The right panels in the same figure show that the adoption of
stellar models with overshooting ($\beta$=0.25) yields larger
values for the minimum mass, and consequently brighter magnitudes.
One can estimate $\sim$ 3.3$M_{\odot}$ at $M_V\sim -$1.7 and
$\sim$ 4.0$M_{\odot}$ at $M_V\sim -$2.4 mag, with $Z$=0.004 and
0.08, respectively. Note that, since at such metal contents
$M_{fl}<$ 2.3$M_{\odot}$ (see Cassisi, Castellani \& Castellani 1997), all the
massive pulsators underwent a quiet ignition of central
He-burning.

Figure 6 shows  that the predicted period-luminosity distribution
of these more metal-rich pulsators appears in good agreement with
the edges of the instability strip at $Z$=0.0004 (solid lines);
the only difference is that the faintest magnitudes are
significantly brighter than the predicted value for $Z$=0.0004
(upper arrow in each panel), with the shortest periods increasing
well above  the $Z$=0.0004 value  (log$P_F=-$0.3, see Fig. 4).
Moreover, since the luminosity level of RR Lyrae stars (dotted
lines) is even fainter than at $Z$=0.0004 (lower arrow in each
panel), the difference in magnitude between the faintest massive
pulsators and RR Lyrae variables becomes larger than $\sim$ 2.0
mag (no overshooting) or $\sim$ 3.0 mag ($\beta$=0.25), against
the value of $\sim$ 1 mag at $Z$=0.0004.

\begin{figure}
\includegraphics[width=10cm]{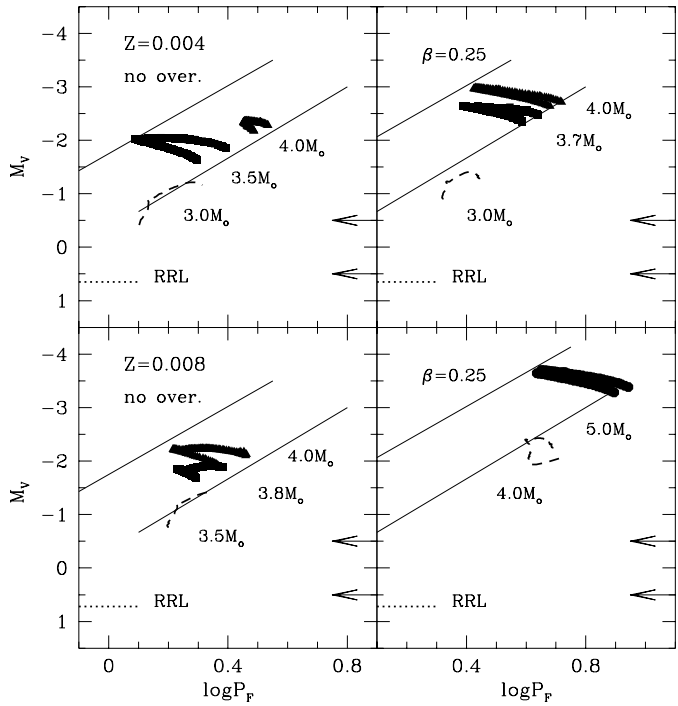}
\caption{As in Fig. 4, but with $Z$=0.004 and $Z$=0.008 and under
different assumptions on the efficiency of overshooting. The solid
lines are the same as in Fig. 4. Note that $M_{V,LE}$ is now
brighter than the value predicted at $Z$=0.0004 (upper arrow in
each panel), while  RR Lyrae stars (dotted line) are slightly
fainter (lower arrow in each panel). Moreover, the shortest period
of massive pulsators ranges from log$P_F\sim$ 0.10 ($Z$=0.004, no
overshooting) to $\sim$ 0.50 ($Z$=0.008, $\beta$=0.25), which are
significantly longer than the value (log$P_F\sim -$0.3) at
$Z$=0.0004. }
\end{figure}

As for mass loss, we have already presented evidence that for
structures with $M$ in the range 1-2 $M_{\odot}$ the post He-flash
evolution is little affected by mass-loss during the RGB phase.
The effects of mass loss in more massive structures which quietly
ignited He-burning ($\sim$ 3-4$M_{\odot}$) has been already
exhaustively discussed in the literature (see, e.g., Bertelli et
al 1989) and one finds that in a metal poor 4$M_{\odot}$ model
even a huge amount of mass loss ($\sim$ 0.13$M_{\odot}$) decreases
the luminosity of the He burning blue loop by less than 0.1 mag
(Castellani et al 2003), with little effect on the predicted
pulsational scenario.

Figure 7 summarizes the results of the whole investigation,
showing the predicted limiting magnitude for massive pulsators
together with the magnitude of RR Lyrae stars, both as a function of
the assumed metallicity over the range $Z$=0.0002 to 0.008. As a
conclusion, we predict that, over this metallicity range, short
period ($P\le$ 3 days) central He-burning pulsators more massive
than RR Lyrae stars populate a common instability strip in the
$M_V$-log$P$ plane, independently of their metal content and of
the occurrence of the He flash. The only discriminating parameters
appear  to be the faintest magnitude and the shortest period, which
become fainter and shorter, respectively, as metal
content decreases.

\begin{figure}
\includegraphics[width=10cm]{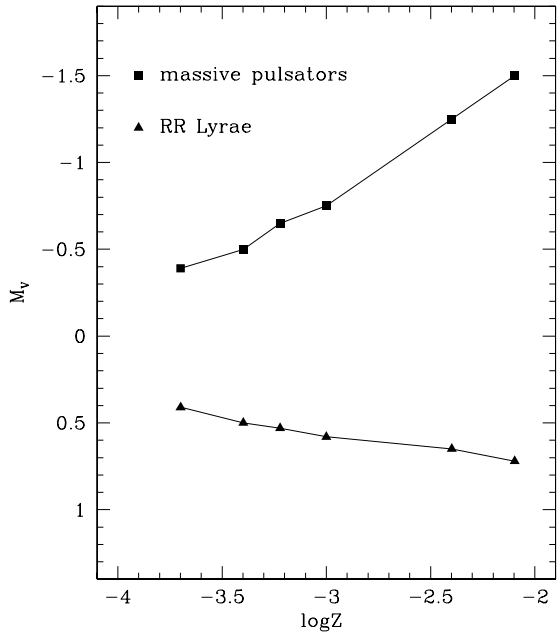}
\caption{The predicted limiting magnitude for massive pulsator
(squares) and the magnitude of ZAHB RR Lyrae variables (triangles)
as a function of the star metallicity.}

\end{figure}

In this respect, present results fully support the suggestion by
Cordier et al (2003) who explain the occurrence of faint short
period Cepheids in the Small Magellanic Cloud as an effect of a
decreased metallicity. Here we only add that inspection of
evolutionary features for metal deficient stars (see Cassisi et al. 1997) discloses that in the extreme case
$Z$= 10$^{-5}$ one expects pulsators with a continuous
distribution of masses, with a lower mass limit at
$\sim$1.0$M_{\odot}$, $M_V\sim$ 0.1 mag.

\section{Comparison with observations}

We can now compare theoretical predictions with the sample of
variables observed in Leo A, as given by D02 but with
fundamentalised periods for FO candidates . Figure 8 shows the
$M_V$-log$P_F$ data for variables with high quality light curves,
assuming a visual distance modulus $\mu_V$=24.6 mag as derived
from the mean magnitude $<V>$=25.1 mag of the observed RR Lyrae
stars and the predicted absolute value of 0.5 mag for $Z$=0.0004
(see Fig. 4). The solid lines depicting the instability strip in
Fig. 4 are repeated here. One finds that predictions and
observations appear in  reasonable agreement, both for the
instability edges as well as for the evidence that bright
pulsators show a limiting magnitude of $\sim -$0.5 mag.

\begin{figure}
\includegraphics[width=10cm]{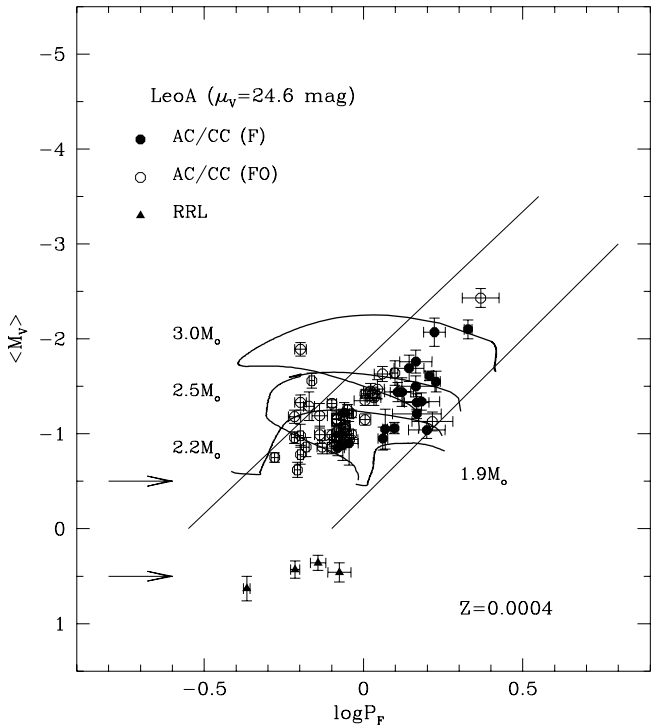}
\caption{Observed Cepheids (dots) and RR Lyrae stars (triangles)
in Leo A as compared with the predicted edges of the instability
strip in the $M_V$-log$P_F$ plane (solid lines). The period of
first-overtone pulsators is fundamentalised. The arrows depict the
theoretically predicted magnitude of the faintest Cepheids
($M_{V,LE}\sim -$0.5 mag) and of RR Lyrae stars ($M_V$(RRL) $\sim$
0.5 mag) at the Leo A metal content ($Z$=0.0004). Selected
evolutionary tracks are also drawn.}

\end{figure}

The comparison with selected evolutionary tracks shows that
Cepheids in Leo A should have masses from $\sim$ 1.9 to
3$M_{\odot}$, where the lower limit is the theoretical prediction
for the occurrence of pulsators, whereas the upper limit is only
indicating the lack of more massive evolving stars. Thus the 
Leo A variables beautifully confirm that the distribution in the
$M_V$-log$P$ diagram of massive pulsators which experienced the He
flash is the natural extension of the distribution of more massive
pulsators characterized by a quiet He ignition, at least for
masses $M\le$ 4.0$M_{\odot}$.

As a relevant point, one finds that the discussed dependence on
the metallicity appears supported by observational data. A
reasonable agreement with theory is indeed obtained for Sextans A,
which is one of the lowest metallicity ($Z\sim$ 0.001) galaxies
with observed Classical Cepheids (see Dolphin et al. 2003).
Adopting the visual distance modulus provided by the authors, we
show in Fig. 9 that the absolute magnitude of the variables ranges
from $M_V\sim -$2.5 mag to $\sim -$0.75 mag, the latter being the
limit predicted by theory for the metallicity of the galaxy. As
shown in the figure, in this case one predicts a minimum mass of
the order of  $\sim$ 2.5$M_{\odot}$, larger than  inferred from the
variables in Leo A, thus suggesting that all the Cepheids in
Sextans A had a quiet He-burning ignition.

\begin{figure}
\includegraphics[width=10cm]{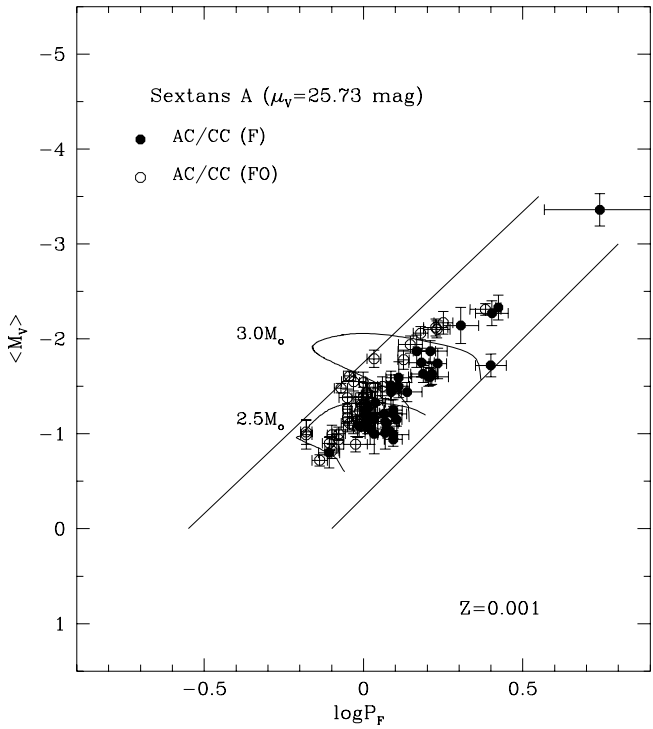}
\caption{As in Fig. 8, but with Classical Cepheids in Sextans A
and  for $Z$=0.001.}
\end{figure}

Before closing this section, let us show in Fig. 10 the comparison
between our predictions with $Z$=0.0004 and data for
well-recognized ACs in dwarf spheroidal galaxies, using the
distance moduli provided by the various authors (see Marconi et al.
 2004 for references). As a whole, one finds a
reasonable agreement as far as the edges of the instability strip
are concerned. However, if the distance moduli provided in the
literature are reliable, several variables appear
significantly fainter than the predicted limiting magnitude at
$Z$=0.0004 ($\sim -$0.5 mag, see arrow), an occurrence which
cannot be explained in terms of smaller mass at such a metal
content (see Fig. 4). Moreover, the study by Marconi et al. (2004) 
does suggest values as low as $\sim$
1.25$M_{\odot}$,  as inferred  by the analysis of intrinsic colors
and visual amplitudes. According to the results presented in the
present paper, namely the decrease of the minimum mass and
luminosity when decreasing the metal content, the faintest ACs
plotted in Fig. 10 suggest the occurrence of variables with metal
contents lower than $Z$=0.0004.

\begin{figure}
\includegraphics[width=10cm]{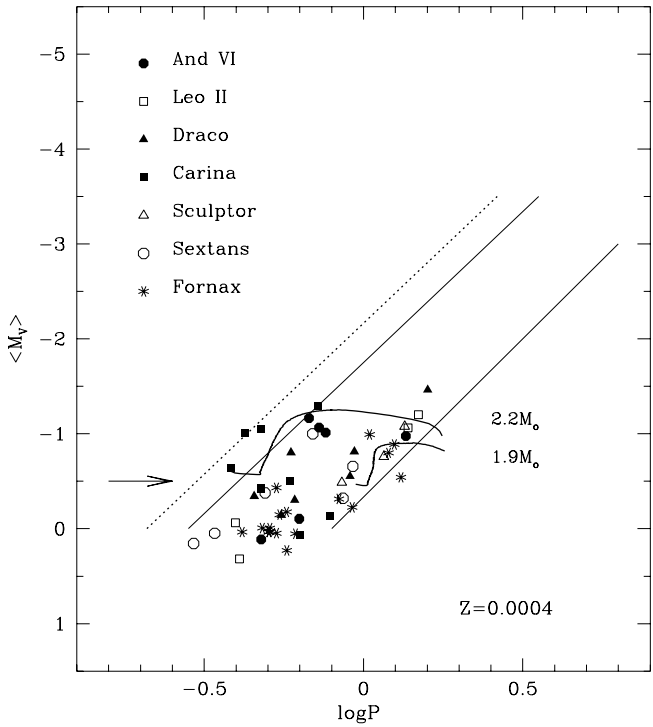}
\caption{Observed Anomalous Cepheids in dwarf spheroidal galaxies
as compared with the predicted edges of the instability strip in
the $M_V$-log$P$ plane. The solid lines refer to fundamental
 edges and are the same as in Fig. 4. In order to account for the
occurrence of first-overtone pulsators, the blue limit is shifted
by $\delta$log$P=-$0.13 (dotted line). The arrow depicts the
theoretically predicted magnitude of the faintest Cepheids
($M_{V,LE}\sim -$0.5 mag) at $Z$=0.0004. The selected evolutionary
tracks are drawn with their fundamental period.}
\end{figure}

\section{Concluding remarks}

In the current literature, the term ``Anomalous'' Cepheids is
widely adopted to indicate variables brighter than RR Lyrae stars
and with periods $P\le$ 2 days, as repeatedly observed in dwarf
spheroidal galaxies of the Local Group and, more in general, in
metal-deficient stellar systems hosting not-too-old stellar
populations. However, in a recent study (Dolphin et al. 2002) it
has been suggested that variables with such features, as observed
in the metal-poor dwarf irregular  galaxy Leo A, can represent the
natural extension to low metallicity of the sequence of Classical
Cepheids, rather than be classified as Anomalous Cepheids.

Starting from such a suggestion, we have studied the
theoretical scenario for $Z$=0.0004 to 0.008 central He-burning
pulsators with mass $\le 4M_{\odot}$, as based on updated
pulsation and evolution models. On this ground, the predicted
edges of the pulsation region are derived, suggesting that $P\le$ 3
days central He-burning pulsators more massive than RR Lyrae stars
populate an almost unique instability strip, independently
of whether they ignited He-burning quietly or flashing, and of
their metal content. The very discriminating parameter is the
minimum mass for the occurrence of these bright
variables, and consequently their faintest magnitude and shortest period
which becomes brighter and longer when moving from
$Z$=0.0004 to 0.008.

 The comparison of predictions with observed variables in Leo
A ($Z\sim$ 0.0004) and Sextans A ($Z\sim$ 0.001) discloses a quite fair agreement, either for what concerns
the distribution of the variables in the $M_V$-log$P$ plane and
the limiting magnitudes and periods. On this basis, present
results {\it do} confirm that the so-called "Anomalous" Cepheids
are the natural extension of the "Classical" Cepheids to lower
metallicities, where the evolution of less massive stars intersects
the pulsation region, producing pulsators with
fainter
magnitudes and shorter periods.

As a conclusion, not surprisingly one finds that going outside our
own Galaxy  the current classification of bright variables tends
to be rather misleading. This because the classical dichotomy
observed in the Milky Way, old and metal-poor stellar populations
against young and metal-rich ones, is vanishing. The problem we
are dealing with is indeed caused by the extragalactic evidence of
rather young but metal-poor populations with their pulsating
instabilities. Following early discussions (see, e.g. Castellani
1986), there is now an increasing consensus in extending the term
of Population II to all metal poor populations, independently of
their age. In this context, there is no reason to still call
``Anomalous'' the massive bright pulsators observed in metal
deficient stellar systems. Wishing to retain the term ``Cepheids''
for pulsators brighter than RR Lyrae stars, and considering that
the only distinctive factor is the pulsator mass, it would be
possibly better to make clear the evolutionary status of the
variables, using the term ``HB Cepheids'' (HBC) for low mass
Cepheids with ZAHB progenitors (Population II Cepheids in the
current nomenclature), whereas the so called ``Anomalous'' should
be simply regarded as bona fide metal-poor, low-mass Classical
Cepheids.

Finally, even though detailed synthetic simulations are needed to
account for the star evolutionary lifetimes, we suggest that the
observed faintest Cepheids could yield a straightforward distance
estimate to the hosting stellar system, provided that no
significative metallicity dispersion is present, the metal content
is well-known and no completeness problem is affecting the Cepheid
sample. We should add that, since the RR Lyrae luminosity
decreases with increasing metallicity, the  observed
difference in magnitude between the faintest Cepheids and RR Lyrae
stars with similar metal content is predicted to be a function of
the metal content, increasing from $\sim$ 1 mag ($Z$=0.0004) to
$\sim$ 2.2 mag and 3.2 mag (at $Z$=0.008 without overshooting and
with $\beta$=0.25, respectively), thus providing a useful metal
content indicator.

\section{Acknowledgments}

We warmly thank Giuseppe Bono and Steven Shore for useful suggestions and
a careful reading of the manuscript.   We also thank an anonymous referee
for his/her useful comments which have considerably improved the paper.
Financial support for this work was provided by the scientific project
``Stellar Populations in Local Group Galaxies'' from MIUR-Cofin 2002
(P.I> Monica Tosi).
 Model computations made use of resources granted by the
``Consorzio di Ricerca del Gran Sasso'', according to Project 6
``Calcolo Evoluto e sue Applicazioni (RSV6) - Cluster C11/B''

{}

\end{document}